# Solar-Blind QKD over Simplified Short-Range FSO Link

**Florian Honz[1], Michael Hentschel[1], Philip Walther[2], Hannes Hübel[1], and Bernhard Schrenk[1]**

[1]AIT Austrian Institute of Technology, Center for Digital Safety & Security / Security and Communication Technologies, 1210 Vienna, Austria.
[2]University of Vienna, Faculty of Physics, Vienna Center for Quantum Science and Technology (VCQ), Boltzmanngasse 5, 1090 Vienna, Austria.
Author e-mail address: florian.honz@ait.ac.at

**Abstract:** We demonstrate QKD and data communication over an out-door free-space link where large-core fiber substitutes active alignment. We further prove E-band QKD as stable and robust under full daylight, despite the loss of spatial filtering.  © 2024 The Author(s)

## 1. Introduction

The rapid progress in the development of quantum computers poses a threat to the currently employed asymmetric security protocols, making them vulnerable to deciphering by eavesdroppers. One way to secure our communications against this threat is the widespread adoption of quantum key distribution (QKD), which guarantees the secure exchange of cryptographic keys by the laws of nature. Whereas QKD systems have reached the technological maturity for deployment in optical fiber networks on a commercial basis, free-space optical (FSO) QKD systems have not reached the same level yet – despite the need of an optical continuum to support QKD in fiber-scarce environments. This immaturity inherent to FSO-based QKD is a result of (*i*) the need for active alignment of the receiver and transmitter stations [1-5], (*ii*) the size of the used telescopes to ensure a good coupling efficiency [1-3], and (*iii*) the detrimental impact of solar irradiance on the quantum signal [3-5]. Even though recent demonstrations of terrestrial QKD systems have shown QBERs as low as 0.5% with key-rates in the kb/s range despite being operated in full daylight [1], the excessive use of spatial and spectral filtering induces a high degree of system complexity, which ultimately hinders the widespread deployment of free-space terrestrial QKD links.

In this work, we propose a simplified setup for free-space QKD based on large-core fiber optics, without large telescopes and active beam steering. Despite the poor spatial filtering, we accomplish solar-blind operation through E-band transmission, using cost-effective CWDM technology. We demonstrate raw-key generation well below the QBER limit at three channels within the atmospheric water absorption windows, yielding a raw-key rate of 3.7 kb/s at a 7.9% QBER at 1410 nm. We further prove co-existence with a data channel with carrier-grade (0 dBm) launch.

## 2. Large-Core Free-Space QKD with Solar-Blind Operation

Single-mode fiber coupling poses an enormous challenge for FSO deployments. Instead of relying on ITU-T G.652 compatible telecom single-mode fiber (SMF) at the receiver terminal, we propose the use of a large-core fiber. As we have demonstrated recently for classical FSO communication, these fibers can significantly relax the alignment tolerance while enabling a fully-passive FSO deployment [6]. Here, we evaluate three different fibers for the receiver in our short-range free-space optical (FSO) link: (*i*) a SMF with a core diameter of 8.2 µm, (*ii*) a step-index multi-mode fiber (MMF) with a 25-µm core diameter, and (*iii*) a graded-index OM4 MMF with a 50-µm core (Fig. 1a). We used commercially available fiber collimators intended for beams with a diameter of ~8 mm diameter at the transmitter and receiver terminals. Their alignment is manually set and frozen for the duration of the experiment.

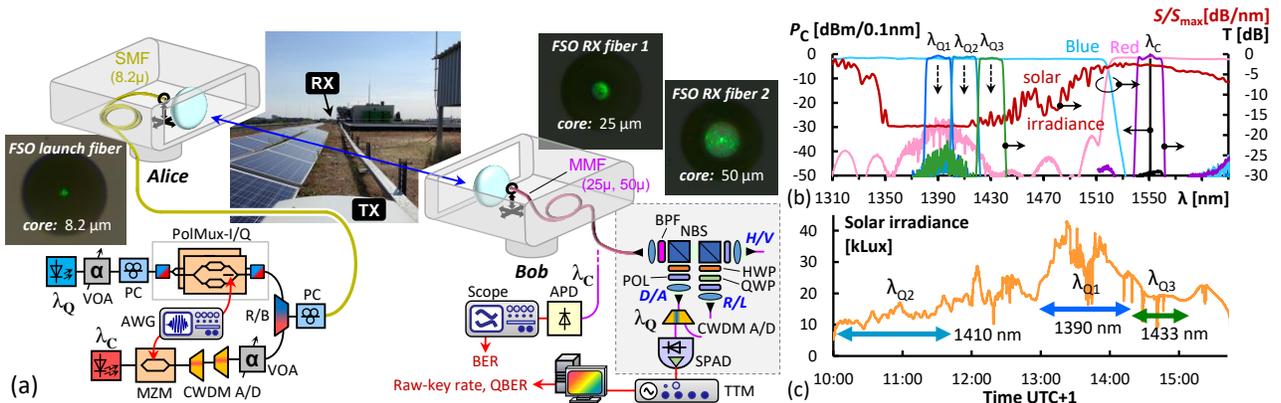

Fig. 1. (a) Experimental setup. (b) Optical spectra, filter transmission $T$ and solar background $S$. (c) Solar irradiance during the QKD evaluation.

The use of large-core fibers stands in a trade-off with the spatial filtering capability of the FSO system, making the QKD link vulnerable to sunlight. Towards this direction, we build on the asymmetry in distance between the short-range (63 m) FSO link and the thickness of the atmosphere. The impact of solar irradiance can be mitigated by spectrally allocating the quantum channel within the water region at the E-band. A long-term characterization of the solar background coupled to a 50-µm core fiber, as acquired in Fig. 2 proves this point with a single-photon spectrometer (10% efficiency, 12 dB roof-top feeder fiber and grating insertion loss before SPAD). The water vapor in the atmosphere creates a wide spectral notch at ~1400 nm, spanning over 4 CWDM channels where we primarily see dark counts at <26 dBHz/nm.

Figure 1a presents the experimental setup to accomplish QKD within this notch. The quantum transmitter is sourced at $\lambda_{Qi}$ = {1390, 1410, 1433 nm}, as highlighted in Fig. 1b together with the respective CWDM windows and the solar influx $S$. We implement a polarization-based BB84 protocol by encoding the states in the right-/left-circular (**R/L**) and anti-/diagonal (**A/D**) bases, using a polarization-multiplexed (PolMux) *I/Q* modulator [7]. Since these states only differ by the relative phase between the two principal horizontal (**H**) and vertical (**V**) polarization states, modulation of the phase section within the *I/Q* modulator via an arbitrary wave form generator (AWG) is sufficient to generate all four states. This phase electrode supports an electro-optic bandwidth of 0.92 GHz, which suits for QKD operation at $R_Q$ = 0.5 Gbaud. A polarization controller (PC) at the output of the PolMux *I/Q* modulator compensates the polarization drift along the fiber-based feeder to and from the FSO terminals. A variable optical attenuator (VOA) within the transmitter sets the mean photon level $\mu_Q$ to 0.1 photons/symbol at the transmitter output.

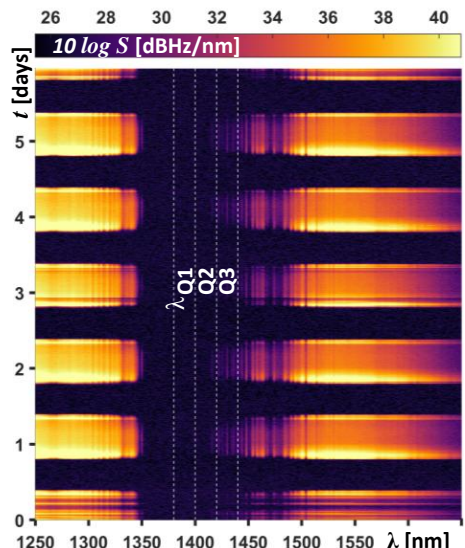

Fig. 2. Single-photon spectrum of solar background.

The robustness to co-existence with data transmission was proven by appending a C-band channel at $\lambda_C$ = 1547.72 nm modulated by on-off keyed (OOK) data at $R_C$ = 1 Gb/s, using a Mach-Zehnder modulator (MZM). We employed two CWDM add/drop filters at 1550 nm to strip off the far-reaching spontaneous emission tails of the classical channel and set the transmit power to 0 dBm. After multiplexing quantum and classical channel with a red/blue (R/B) filter, both signals were fed to the roof-top through a 200-m SMF as launch fiber for the FSO link. At the receiving side, the signal that is coupled to the large-core fiber was relayed by a 1-m patchcord to a polarimeter that is equipped with free-space polarizers to analyze the signal in the **R/L**, **A/D** and **H/V** bases. An InGaAs SPAD then detects the signal. It is characterized by a dark count rate of 300 cts/s, a quantum efficiency of 10 % at 1550 nm and a dead time of 25 µs. The SPAD counts were then recorded by a TTM module and submitted to temporal filtering at 50% of the symbol width before a real-time estimation of raw-key rate and QBER were conducted. The classical channel was detected by a commercial APD receiver and evaluated by means of BER estimation.

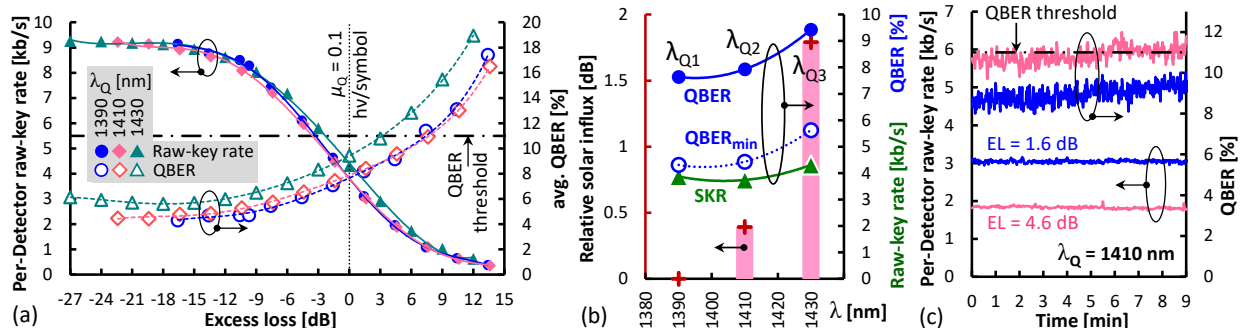

Fig. 3. (a) QKD performance as function of excess link loss and (b) comparison for the three chosen E-band CWDM channels. (c) QKD stability.

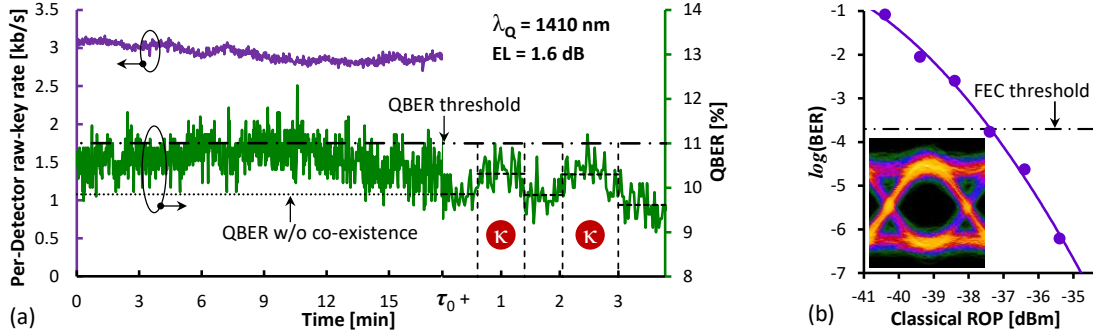

Fig. 4. (a) QKD performance under co-existence. (b) BER performance for classical data transmission, including its eye diagram.

## 3. QKD Performance over Large-Core Free-Space Link

The spectral layout of the QKD link is presented in Fig. 1b. The QKD channel ($\lambda_{Qi}$) inside the water region around 1400 nm is cleaned from out-of-band sunlight background noise by a cascade of a 50-nm free-space bandpass filter (BPF) at the polarimeter and multi-mode CWDM add/drop filter before the SPAD. While the former provides good suppression over the entire responsivity window of the SPAD, the latter features a lower insertion loss. This filter cascade enabled us to reduce the background noise to a level below the dark count level of the SPAD for $\lambda_{Q1}$ and $\lambda_{Q2}$, despite the poor spatial filtering at the FSO receiver due to coupling to large-core fiber. For $\lambda_{Q3}$ at the border of the water vapor line, we saw an elevated in-band noise floor at 590 cts/s, which is about twice the dark count rate.

The performance evaluation was conducted for all three quantum channels $\lambda_{Qi}$ and different large-core fibers on a single day in autumn, characterized by light clouds in the morning and late afternoon, and full sunshine during noon. Figure 1c reports the solar irradiance together with the evaluation periods of the three CWDM channels.

*Evaluation of large-core fibers:* When using *SMF* as the receiving fiber, we were not able to obtain a stable link loss of <24 dB without active beam alignment. This renders SMF as unsuitable for a low-complexity FSO setup. For the *50-µm OM4 MMF* we were able to reduce the FSO link loss to 7 dB. However, the QBER of 19% was clearly above the threshold of 11% for generating a secure key. This is attributed to depolarization occurring in combination with the high number of spatial modes excited within the MMF, resulting in a non-uniform polarization evolution of the guided modes despite a relatively short length of MMF [8]. For the *25-µm MMF* we accomplished a FSO transmission of -17.8 dB while at the same time reaching QBER values that fall within a 1% penalty compared to a fully SMF-based reference link without FSO segment. This renders the 25-µm MMF a suitable candidate for the large-core FSO link: There was no need to re-align the FSO coupling during the entire measurement campaign.

*Evaluation of optimal wavelength:* Figure 3a reports the average QBER and per-detector raw-key rate for each quantum channel over the tolerable optical excess loss (EL) of the FSO channel, relative to the nominal mean photon level of $\mu_Q = 0.1$ photons/symbol. As can be seen, the results suggest 1390 and 1410 nm as the preferred choice, reaching a per-detector raw-key rates of 3.7 kb/s at a QBER of 7.9% (●,♦). We can accommodate an additional optical budget of up to 7.6 dB before surpassing the QBER threshold of 11%, which effectively translates to a comfortable link margin for the passive FSO layout. The QBER at 1430 nm is slightly increased to 9.4% (▲), which is attributed to the spectral roll-off of the water vapor absorption line (Fig. 2). Figure 3b summarizes the channel performance, together with the relative solar influx to the large-core fiber. We then investigated the continuity in performance for two EL settings of 1.6 and 4.6 dB, in order to rule out artifacts specific to the proposed large-core FSO concept, such as sporadic excitation of higher-order modes due to vibrations of atmospheric conditions. Figure 3c reports a stable raw-key rate and QBER for both EL values, without negatively affected blocks in acquired SPAD counts. The slight increase in QBER is explained by the polarization drift along the launch fiber.

*Classical co-existence:* We finally investigated the impact due to simultaneous classical data transmission (Fig. 4a). The QBER increases by ~0.7% in average, as it is evidenced for block-wise (45 s) activation (κ) of the data channel. The BER performance of the classical channel is reported in

Fig. 4b. The reception sensitivity at the RS(255,239) FEC threshold of $2\times10^{-4}$ is -37.4 dBm. This yields a power margin of >15 dB for the FSO link.

## 5. Conclusion

We have experimentally demonstrated a low-complexity free-space QKD link with alignment-tolerant large-core fiber and simple optical FSO setup. Stable operation without significant depolarization due to excitation of unfavorable higher-order modes has been proven for polarization-encoded BB84 QKD, rendering the 25-µm core fiber as suitable fiber medium. The lack of spatial filtering has been addressed through the spectral allocation of the quantum channel in the water region of the E-band, resulting in solar-blind operation at a raw-key rate of 3.7 kb/s and a QBER of 7.9% for the 1390 and 1410-nm CWDM channels, even in co-existence with a 1-Gb/s C-band data channel. This enables a considerable simplification of short-range terrestrial FSO QKD operated at full daylight.

Acknowledgement: This work was supported by the Digital European Program under Project number No 101091642.